\documentclass[journal=mamobx,manuscript=communication]{achemso}

\usepackage[version=3]{mhchem} 
\usepackage{subfigure}
\usepackage{color}

\author{Ting Ge}
\affiliation{Department of Physics and Astronomy, Johns Hopkins University, Baltimore, MD 21218 USA}
\author{Gary S. Grest}
\affiliation{Sandia National Laboratories, Albuquerque, NM 87185 USA}
\author{Mark O. Robbins}
\email{mr@jhu.edu}
\affiliation{Department of Physics and Astronomy, Johns Hopkins University, Baltimore, MD 21218 USA}
\title[\texttt{achemso} demonstration]
{Structure and Strength at Immiscible Polymer Interfaces}

\begin{document}
\begin{abstract}
Thermal welding of polymer-polymer interfaces is important for integrating polymeric elements into devices. When two different polymers are joined, the strength of the weld depends critically on the degree of immiscibility. We perform large-scale molecular dynamics simulations of the structure-strength relation at immiscible polymer interfaces. Our simulations show that immiscibility arrests interdiffusion and limits the equilibrium interfacial width. Even for weakly immiscible films, the narrow interface is unable to transfer stress upon deformation as effectively as the bulk material, and chain pullout at the interface becomes the dominant failure mechanism. This greatly reduces the interfacial strength. The weak response of immiscible interfaces is shown to arise from an insufficient density of entanglements across the interface. We demonstrate that there is a threshold interfacial width below which no significant entanglements can form between opposite sides to strengthen the interface.   
\end{abstract}

\begin{tocentry}
\includegraphics{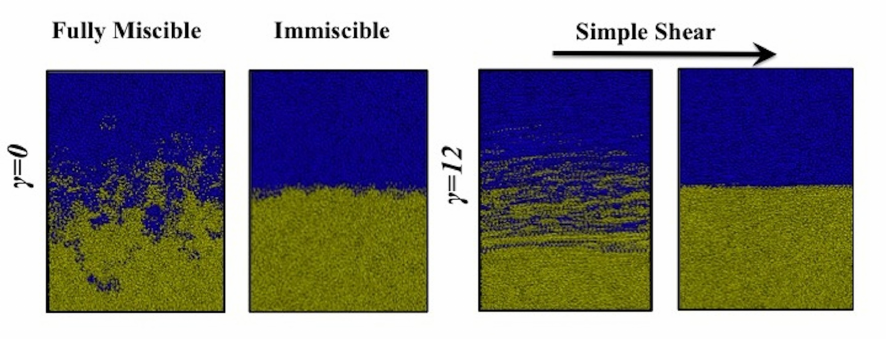}
\end{tocentry}
Disparate polymers usually do not mix well\cite{helfand71,helfand72,rubinstein03,robeson07}. Since even a small energy penalty associated with contact of different constituent monomers is amplified by the high degree of polymerization, the enthalpic contribution to the free energy often dominates over the entropy gain due to partial mixing. As a result, an equilibrium interface of limited width forms between immiscible polymers.  This type of interface exists in numerous applications of polymer blends\cite{robeson07} and exhibits low mechanical strength during large deformation and fracture\cite{wool95,jone99}.  An understanding of the molecular origin of this weakness may aid development of novel techniques for reinforcing immiscible polymer interfaces.   

Computer simulations access molecular details that are difficult to observe in experiments and thus provide unique insight into interfacial structure and mechanical processes. In particular, recently developed algorithms\cite{everaers04,kroger05,tzoumanekas06} have enabled simulations to track entanglements on a microscopic level. The entanglement density can be directly related to the viscoelastic response of high molecular weight polymer melts.\cite{degennes71,doi88,rubinstein03} Experiments have suggested that entanglements strongly affect the mechanical properties of interfaces between glassy polymers,\cite{wool95, schnell98,schnell99,brown01,creton02,cole03,boiko12b}
and many theoretical models also assume that entanglements play a critical role.\cite{degennes89,wool95,benkoski02,silvestri03}   

In this Letter, we present results from large-scale molecular dynamics (MD) simulations of the interdiffusion between highly entangled immiscible polymers. As the degree of immiscibility increases, the equilibrium interfacial width decreases and is reached at an earlier interdiffusion time $t$. The interfacial strength of an immiscible interface is always lower than that of a fully miscible interface at the same $t$ and saturates below the bulk strength in equilibrium. Immiscible interfaces are not able to transfer stress effectively because chains can pull out from the opposing surface, while failure of bulk systems requires chain scission. We use the Primitive Path Analysis (PPA) algorithm \cite{everaers04,hoy07b} to identify entanglements and correlate them with mechanical response. Entanglement densities are greatly reduced at immiscible interfaces relative to bulk values and we find that no entanglements form across the interface for interdiffusion depths below a threshold value.

\begin{figure}[htb]
\includegraphics[width=0.45\textwidth]{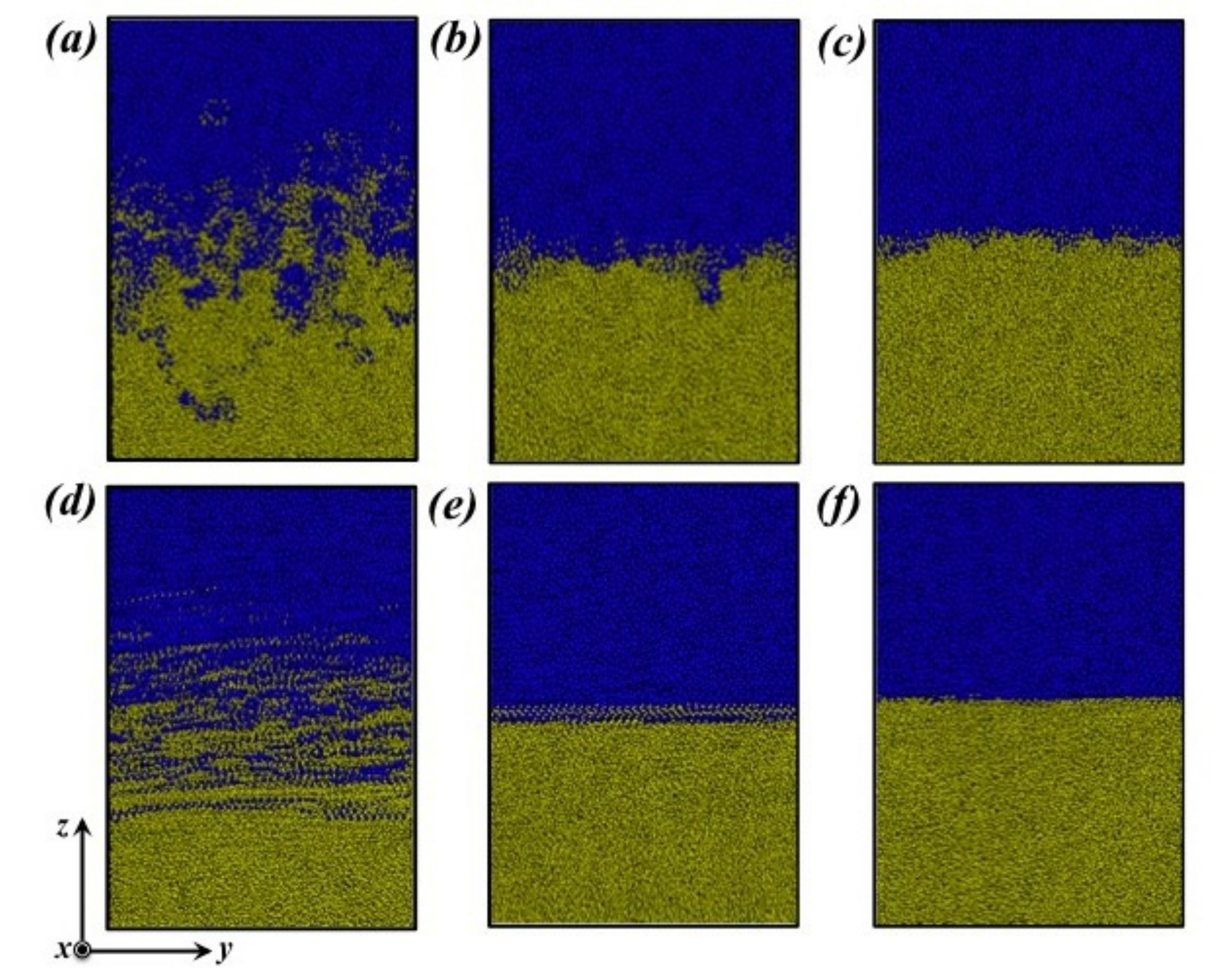}
\caption{Snapshots of the interface between thin polymer films of
type 1 (yellow) and 2 (blue) at $T=0.2u_0/k_B$ before deformation (top) and at a large shear strain $\gamma=12$ (bottom).
Snapshots (a,d) are for a fully miscible interface ($\tilde{\epsilon}_{12}=1.0$) at interdiffusion time $t=5 M\tau$, (b,e) and (c,f) show
equilibrium states for $\tilde{\epsilon}_{12}=0.99$ and $0.95$, respectively.
For clarity only a portion of the sample is shown: $40a$ along the direction of shear (y) and $60a$ in the velocity gradient direction (z).
}
\label{fig:snapshot}
\end{figure}

All of the simulations employed the canonical bead-spring model \cite{kremer90} that captures the properties of linear homopolymers. The van der Waals interactions between like monomers of mass $m$ are modeled using the standard Lennard-Jones potential with interaction strength $u_0$, diameter $a$ and characteristic time $\tau=a\sqrt{m/u_0}$.
To model immiscible films, the interaction strength $u_0$ between unlike monomers was reduced to $\tilde{\epsilon}_{12}u_0 <u_0$. Here we simulated four systems with $\tilde{\epsilon}_{12}=1.0$, $0.99$, $0.98$ and $0.95$.

Chains of length $N=500$ beads were made by coupling nearest-neighbors with an additional potential. Since chain scission plays an essential role in the mechanical tests, the usual unbreakable finitely extensible nonlinear elastic (FENE) potential\cite{kremer90} was replaced by a simple quartic potential with the same equilibrium spacing and a breaking force of $240u_0/a$. This is 100 times higher than the maximum attractive force for the Lennard-Jones potential, which is consistent with experiments and previous simulations \cite{rottler02a,stevens01,ge13}.  Previous work has shown that the entanglement length for this model is $N_e = 85 \pm 7$ and that the mechanical response for $N=500$ is characteristic of highly entangled (large $N$) polymers \cite{rottler02a, rottler02b, rottler03, hoy07, hoy08}. Further simulation details can be found in the Supplemental Information.

Fluid films of each polymer species were equilibrated separately at temperature $T=1.0u_0/k_B$. Each film contains 2.4 million beads in $M=4800$ chains. Periodic boundary conditions were applied along the $x$- and $y$-directions with dimensions $L_x=700a$ and $L_y=40a$, while featureless walls separated by $L_z=100a$ confined films in the non-periodic $z$- direction. Equilibrated films were placed in contact and allowed to interdiffuse for a time $t$. The system was then quenched rapidly to $T=0.2 u_0/k_B$, which is below the glass temperature $T_g \approx 0.35 u_0/k_B$ \cite{rottler03c}.  To test mechanical strength, shear was applied to the glassy interface in a manner similar to a shear test of a lap joint in experiments \cite{wool95} and recent simulations \cite{ge13}. Interfaces for different $\tilde{\epsilon}_{12}$ before and after shearing are visualized in \ref{fig:snapshot}.

\begin{figure}[htb]
\includegraphics[width=0.45\textwidth]{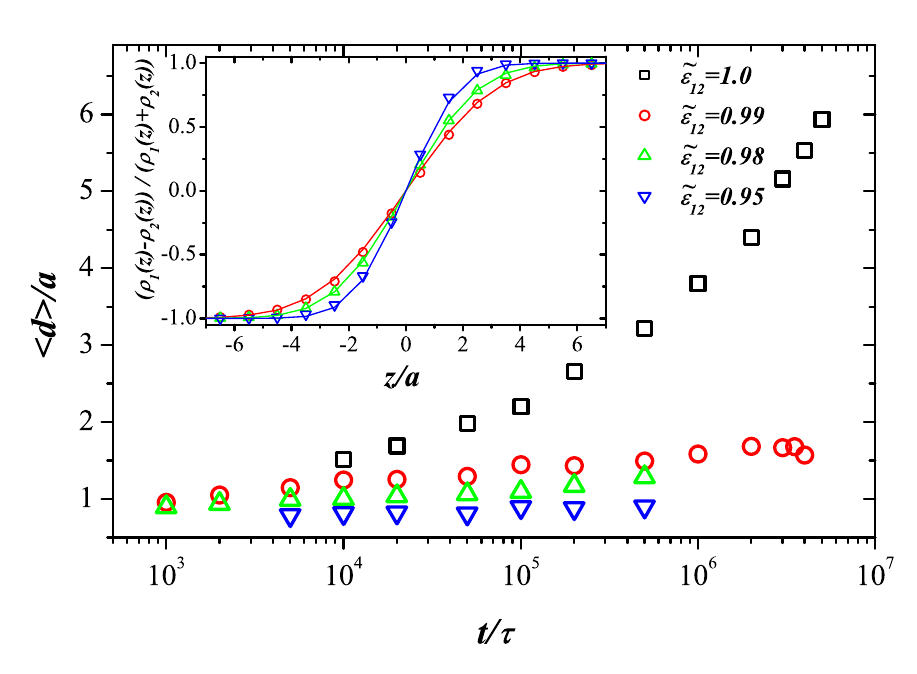}
\caption{Average interdiffusion depth $\left<d\right>$ of monomers across the interface as a function of $t$ for different $\tilde{\epsilon}_{12}$. The inset shows the normalized density difference profile for immiscible interfaces at equilibrium. Solid lines show fits to an error function ${\rm erf}\left(\sqrt{\pi} z/w\right)$.
}
\label{fig:interface}
\end{figure}

\ref{fig:interface} shows the average interdiffusion depth $\left<d\right>$ of monomers across the interface as a function of time $t$ for different $\tilde{\epsilon}_{12}$.
For monomers of type 1 that are below the initial interface ($z=0$),
$\left< d \right> \equiv \int_{0}^{+\infty} (z)\rho_1(z)\,dz /\int_{0}^{+\infty}\rho_1(z)\,dz $, where $\rho_1(z)$ is number density.
For monomers of type 2, the integration range is changed to $z< 0$.
The depth keeps rising with $t$ for $\tilde{\epsilon}_{12}=1.0$, since interdiffusion between fully miscible polymers is a kinetic process which continues indefinitely. In contrast, for $\tilde{\epsilon}_{12}\le 0.99$, $\left<d\right>$ increases slowly with $t$ and reaches a plateau when the entropy gained from mixing is balanced by the energetic penalty. We use the states at $t=4M\tau$, $0.5M\tau$ and $0.5M\tau$ ($1M\tau=10^6\tau$) to represent the equilibrium interface for $\tilde{\epsilon}_{12}=0.99$, $0.98$ and $0.95$, respectively. The corresponding plateau values of $\left<d\right>$ are about $1.6a$, $1.3a$ and $0.9a$. All are much smaller than the width for $\tilde{\epsilon}_{12}=1.0$ at the same times. Separate simulations\cite{pierce11,ge13b} of the self-diffusion of polymer chains with $N=500$ in bulk melts find that the entanglement time $\tau_e\sim10^4\tau$ while the disentanglement time $\tau_d\sim30M\tau$. However, one needs to be careful when comparing the times for interdiffusion with the characteristic times ($\tau_e$ and $\tau_d$) for self-diffusion, since the interdiffusion at early times is found to be dominated by the motion of chain ends\cite{pierce11} and also in this study the interdiffusion is affected by the immiscibility. The reduction of interfacial width due to increasing immiscibility is illustrated by snapshots in \ref{fig:snapshot}(a), (b) and (c). Note that a $1\%$ decrease of $\tilde{\epsilon}_{12}$ from $1.0$ to $0.99$ already leads to a narrow interface with a finite $\left<d\right>$. The sensitivity of interfacial structure to a slight dissimilarity between unlike monomers is well captured by our simulation.

The equilibrium interface width of immiscible polymers is often quantified by the concentration profile\cite{wool95,jone99}.
The inset of \ref{fig:interface}(a) shows $(\rho_1(z)-\rho_2(z))/(\rho_1(z)+\rho_2(z))$.
Solid lines are results from fitting the data points using the error function ${\rm erf}\left(\sqrt{\pi} z/w\right)$.\cite{lacasse98} 
Here $w=4\left<d\right>$ characterizes the equilibrium interfacial width.
Measured values of $w/\left<d\right>$ are consistent with this ratio ($w=6.13\pm0.04a$, $4.98\pm0.04a$ and $3.63\pm0.04a$ for $\tilde{\epsilon}_{12}=0.99$, $0.98$ and $0.95$, respectively).
Helfand and Tagami argued that the width should be twice the radius of gyration
of the chain segments that penetrated across the interface, and that these should have length $1/\chi$ where $\chi$ is the phenomenological Flory interaction parameter.
Then $w\sim 2 \left[ l_K l_0/ 6\chi\right]^{1/2}$, where $l_K=1.77 a$ is the Kuhn length and $l_0=0.96 a$ the bond length.
In our model, $\chi$ scales with $(1-\tilde{\epsilon}_{12})$, but the exact mapping between them is not clear.
While this prevents us from testing Helfand and Tagami's expression, we can use it to estimate that $\chi=0.030$, 0.046 and 0.086 for $\tilde{\epsilon}_{12}=0.99$, 0.98 and 0.95, respectively.

\begin{figure}[htb]
\centering
 \includegraphics[width=0.45\textwidth]{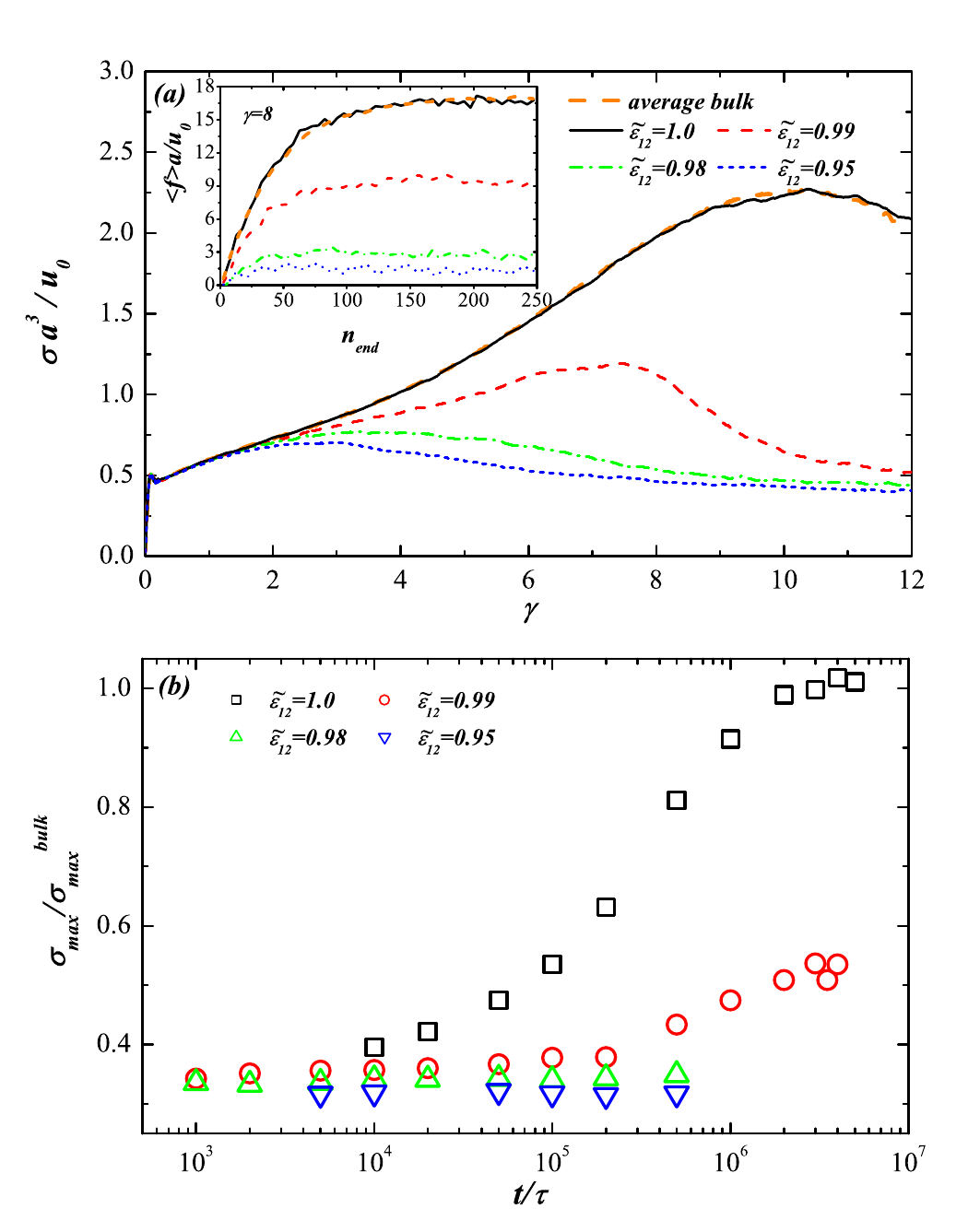}
\caption{(a) Stress-strain curves from shear tests on the fully miscible interface at $t=5M\tau$ and immiscible interfaces at equilibrium. Also shown is the average bulk result. The inset shows the corresponding average bond tension $\left<f\right>$ as a function of the distance in monomers $n_{end}$ from the nearest chain end for $\gamma=8$. (b) The maximum shear stress $\sigma_{max}$ before failure normalized by the average bulk value $\sigma_{max}^{bulk}$ as a function of $t$.   
}
\label{fig:strength}
\end{figure}

\ref{fig:strength}(a) illustrates how the reduced interfacial width changes stress-strain curves from the bulk response, which is the same for both species.
All stress curves show nearly the same initial regimes of linear elastic response, yield and strain hardening as the shear strain $\gamma$ increases. 
For $t > 4M\tau$ the response for $\tilde{\epsilon}_{12}=1.0$ is indistinguishable from the average bulk result even though polymers have diffused by much less than their radius of gyration.\cite{ge13}
As $\tilde{\epsilon}_{12}$ decreases, the stress drops below the bulk response at progressively earlier strains.
Greater immiscibility also lowers the peak stress $\sigma_{max}$ where failure occurs.

As in experiment\cite{wool95}, we use $\sigma_{max}$ to characterize the interfacial strength. \ref{fig:strength}(b) shows $\sigma_{max}$ normalized by the average bulk failure stress $\sigma_{max}^{bulk}$ versus time $t$ for different $\tilde{\epsilon}_{12}$. For $\tilde{\epsilon}_{12}=1.0$, the bulk strength is recovered by approximately $4M\tau$. For $\tilde{\epsilon}_{12}=0.99$, the development of $\sigma_{max}$ is delayed and starts to rise around $0.2M\tau$. Ultimately, it reaches a plateau value that is about one half of $\sigma_{max}^{bulk}$. For $\tilde{\epsilon}_{12}=0.98$ and $0.95$, there is almost no change in $\sigma_{max}$ with the interdiffusion time $t$, and $\sigma_{max}$ remains at a lower value. Similar reductions in the strength of interfaces between two immiscible polymers have been observed in experiments\cite{wool95}. 

Simulations allow us to directly follow the evolution of interfacial structure during shearing and to determine the failure mechanism.
Previous simulations\cite{ge13} revealed that bulk systems fail through chain scission. The same mechanism occurs at long $t$ for $\tilde{\epsilon}_{12}=1.0$,
and the fact that broken bonds are spread uniformly through the sample rather than near the interface confirms that the interface is as strong as the bulk
for $t \ge 4M\tau$. \ref{fig:snapshot}(d) illustrates the distribution of monomers at $\gamma=12$ for this limit. The chain segments that have diffused across the interface become highly oriented during shearing, but have been broken off and continue to shear with the opposing film.

Immiscible interfaces with $\tilde{\epsilon}_{12}=0.95$ and $0.98$ fail through chain pullout at the interface. As illustrated in \ref{fig:snapshot}(f) for $\tilde{\epsilon}_{12}=0.95$, there is a sharp interface at $\gamma=12$ with all chain segments pulled out from the opposing film. The same mechanism is observed for $\tilde{\epsilon}_{12}=1.0$ at short times.
For $\tilde{\epsilon}_{12}=0.99$, the failure mechanism is beginning to crossover from chain pullout to chain scission. However, as shown in \ref{fig:snapshot}(e), only a tiny fraction of monomers remain in the opposite side at $\gamma=12$ and the rest have have been pulled out. Bonds that have broken by $\gamma=12$ are predominantly distributed near the interface, indicating that it is mechanically weaker than the surrounding bulk regions.  

These changes in failure mechanism are directly correlated with the rise in tension along backbone bonds that accompanies strain hardening at large strains.\cite{hoy07}
The inset in \ref{fig:strength}(a) shows the mean bond tension $\left<f\right>$ as a function of the number of bonds $n_{end}$ to the nearest chain end.
Results are shown for $\gamma=8$ where the results for $\bar{\epsilon}_{12}=0.95$ and 0.98 have saturated and the rate of bond breaking is fastest for
$\bar{\epsilon}=0.99$ and $1.0$.
The curves can be fit to $\left<f\right>=f_0 (1-exp\left(-n_{end}/n_{end}^c\right))$ where $f_0$ corresponds to the plateau tension far from ends,
and $n_{end}^c$ is the characteristic distance for tension relaxation near chain ends.
For $\tilde{\epsilon}_{12}=1.0$, the whole distribution of $\left<f\right>$ overlaps with that in the bulk, consistent with the results for the stress-strain behavior.
The length near the end where stress has relaxed, $n_{end}^c$, is only about half $N_e$ and much smaller than the length of chain segments that have diffused across the interface.
While $f_0$ is substantially smaller than the force for chain scission, there is a long tail in the distribution that decays exponentially with a characteristic
decay force equal to $f_0$.\cite{rottler02b}
This allows enough chain scission to produce failure - about 1 in $10^4$ bonds at any time.
As immiscibility increases, the maximum $f_0$ decreases until there is negligible scission.
The value of $n_{end}^c$ also decreases, with $n_{end}^c=42\pm 2$, $33\pm 2$, $21\pm 2$ and $13\pm 3$ for $\tilde{\epsilon}_{12}=1.0$, $0.99$, $0.98$ and $0.95$, respectively.
End segments with length of order $n_{end}^c$ can pullout from their confining tubes.
We find the length in beads $n^*$ of segments that diffuse across the interface is very close to $n_{end}^c$ for
systems that fail by chain pullout:
$33$, 22 and 12 for $\tilde{\epsilon}_{12}=0.99$, 0.98 and 0.95, respectively.
These lengths are obtained using Helfand and Tagami's estimate that $w/2$ corresponds to the radius of gyration of chains of length $n^*$.
Experiments have also observed chain pullout at weak immiscible polymer interfaces.\cite{wool95,schnell99,creton02} 

\begin{figure}[htb]
\includegraphics[width=0.45
\textwidth]{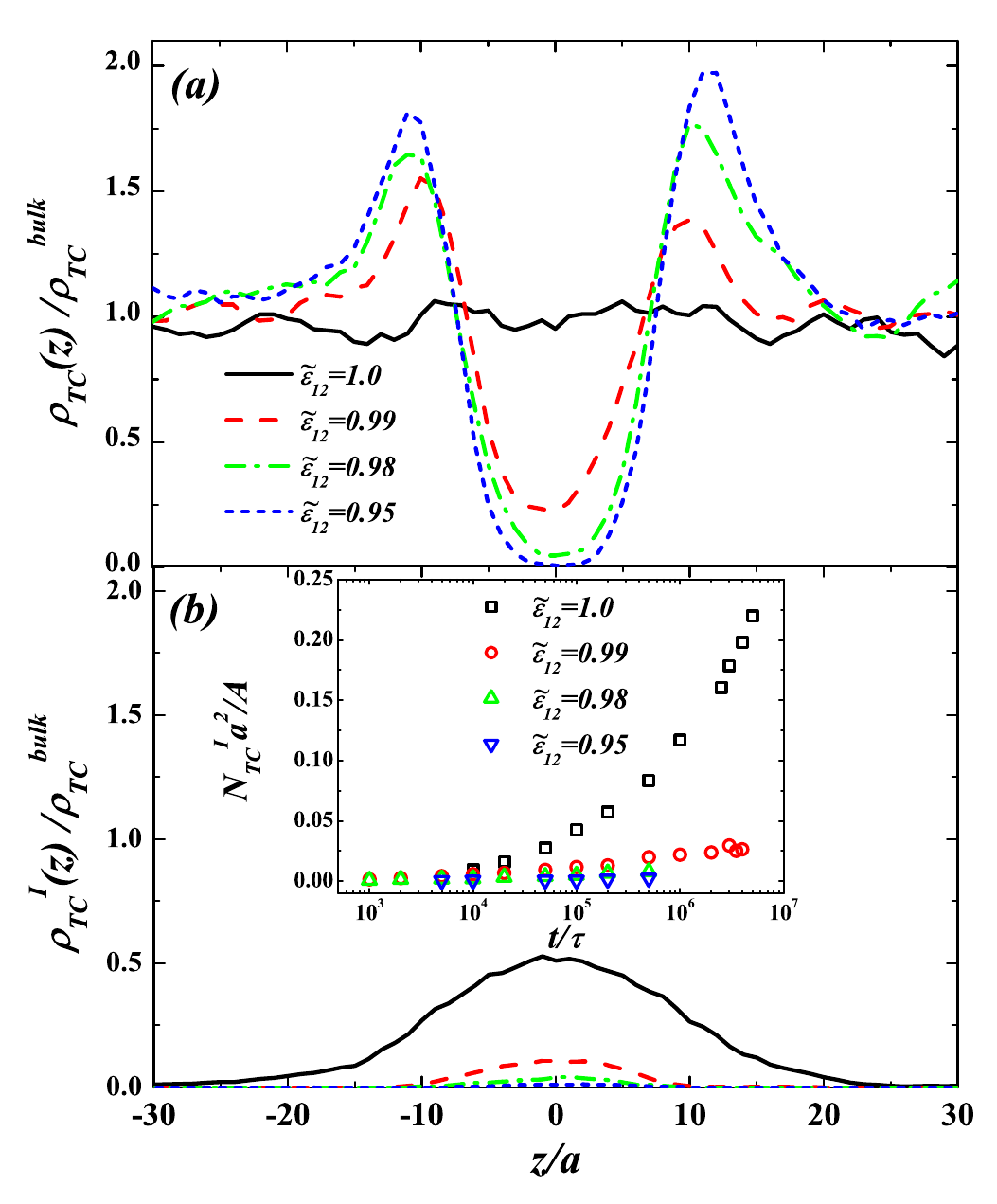}
\caption{ Density profiles of (a) all and (b) interfacial TCs for the same interfaces shown in \ref{fig:strength}. Results are normalized by the bulk density of TCs. The inset of (b) shows the areal density of interfacial TCs, $N_{TC}^I/A$, vs. $t$. 
}
\label{fig:TC}
\end{figure}

Our previous study of miscible interfaces\cite{ge13} showed that entanglements between chains from opposite sides of the interface were required to prevent
chain pullout and lead to chain scission.
One would expect that chain pullout at 
immiscible interfaces also results from a lack of interfacial entanglements.
To test this idea, we tracked entanglements using the PPA, which identifies entanglements as binary contacts between the underlying primitive paths of polymer chains. PPA has provided unique insights into properties of entangled polymer melts\cite{everaers04,kroger05,tzoumanekas06}, because entanglements have remained elusive objects in experimental studies.

In PPA, the primitive paths are revealed by fixing the chain ends and minimizing the chain length without allowing chain crossing. To limit excluded volume effects, the chain diameter is then reduced by a factor of 4 and additional monomers introduced to prevent chain crossing\cite{hoy07b}. Contacts between the resulting primitive paths are counted to determine the number of topological constraints (TCs). We find that the ratio of the density of TCs, $\rho_{TC}$, to the bulk density, $\rho_{TC}^{bulk}$, is insensitive to the procedural details in identifying the TCs. Past studies on bulk polymers have shown that $\rho_{TC}$ is proportional to the entanglement density\cite{everaers04,kroger05,tzoumanekas06,hoy07b}, and we refer to TCs and entanglements interchangeably below.

\ref{fig:TC}(a) shows the profile of $\rho_{TC}(z)/\rho_{TC}^{bulk}$ for the same interfaces shown in \ref{fig:strength}. For $\tilde{\epsilon}_{12}=1.0$, the bulk entanglement density is recovered across the interface at $t \ge 4M\tau$, when the bulk mechanical response has also been recovered. For $\tilde{\epsilon}_{12}<1.0$, the density of entanglements is greatly reduced at the interface. $\rho_{TC}(z)$ is very small near the interface for $\tilde{\epsilon}_{12}=0.99$ and $0.98$ and is essentially non-existent for $\tilde{\epsilon}_{12}=0.95$. This trend correlates with the reduction of interfacial strength as immiscibility increases.

The distributions of TCs for immiscible interfaces exhibit two peaks on either side of the interface. This reflects the anisotropic conformation of chains, which are compressed normal to the free surface before interdiffusion\cite{silberberg82, theodorou88,silberberg88}. Chains with pancake-like conformations near the interface are subject to more TCs. Because immiscibility limits the interdiffusion, these chains cannot relax to their isotropic conformation as in the miscible case. As a result, the peaks in $\rho_{TC}(z)$ at the interface are preserved at equilibrium. Note that the position of the peak in TC density may be shifted by the PPA, which introduces a tension to shorten chain contour lengths that may move TC's towards places with higher density. However the changes in density with time and the total number of interfacial TC's are not sensitive to such shifts.

Interfacial entanglements between chains from two sides are crucial to anchoring chain segments to the opposite side. The distributions of these interfacial entanglements are shown in \ref{fig:TC}(b).
The inset of \ref{fig:TC}(b) shows how immiscibility arrests the formation of interfacial entanglements.
The areal density of interfacial TCs, $N_{TC}^I /A$, is plotted against $t$ for the four values of $\tilde{\epsilon}_{12}$.
For $\tilde{\epsilon}_{12}=1.0$, $N_{TC}^I/A$ continues to increase with interdiffusion time. When chains have formed 2-3 interfacial entanglements, chain pullout is suppressed and bulk strength is achieved\cite{ge13}. 
For $\tilde{\epsilon}_{12}=0.99$, the number of interfacial entanglements is greatly suppressed, while for $\tilde{\epsilon}_{12}=0.98$ and $0.95$, there are almost no interfacial entanglements.
  
\begin{figure}[htb]
\includegraphics[width=0.45\textwidth]{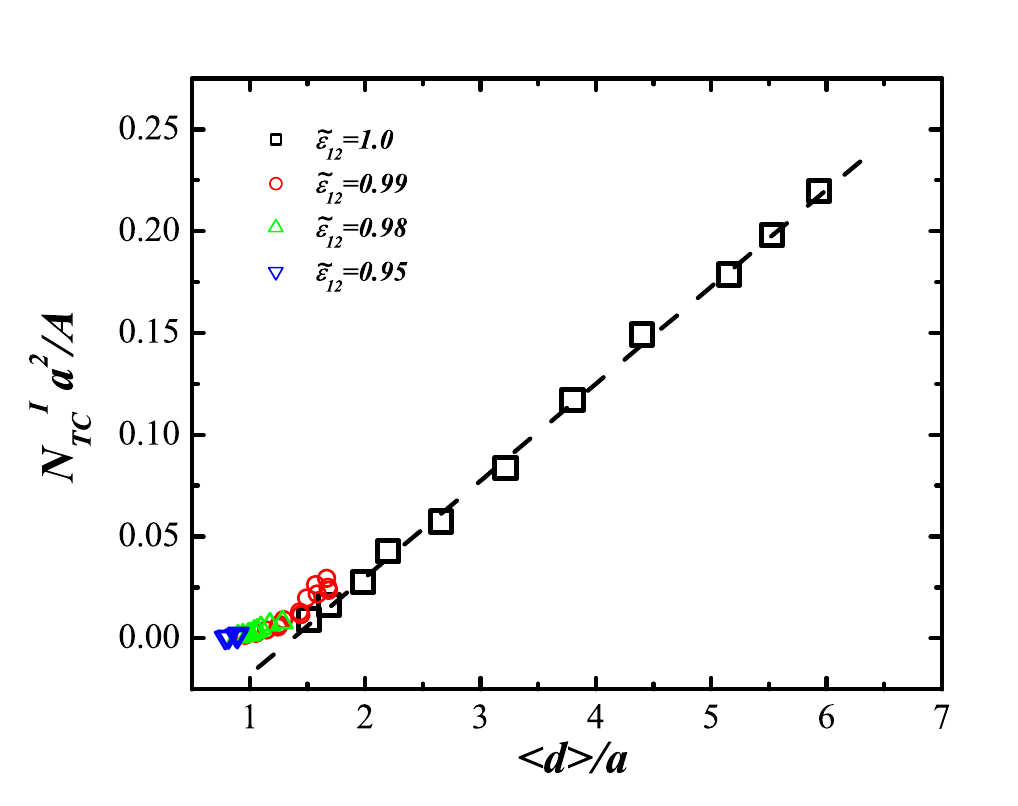}
\caption{The areal density of interfacial TCs, $N_{TC}^I/A$, versus the average interdiffusion depth $\left<d\right>$. Dashed line is the linear fit for the fully miscible case.  
}
\label{fig:ITC}
\end{figure}

Experiments do not provide a direct measurement of entanglements and the interfacial width has often been used as an indirect measure \cite{schnell98,schnell99,brown01,cole03}.
Our simulations allow us to quantify the relation between $N_{TC}/A$ and $\left<d\right>$.
\ref{fig:ITC} shows that results for different $\tilde{\epsilon}_{12}$ are consistent with a common curve.
At large widths, $N_{TC}^I/A$ rises linearly with $\left<d\right>$. We have shown that this agrees with a scaling prediction based on the chain-packing model\cite{ge13}.  The linear region extrapolates to $N_{TC}^I=0$ at $\left<d\right>\sim1.5a$, and the density of entanglements is nearly zero for widths below this threshold value.
For $\tilde{\epsilon}_{12}=0.99$ 
the width rises above this threshold and a slight upturn in entanglement density starts near $\left<d\right>=1.5$.
For less miscible systems $\left<d\right>$ remains below $1.5a$ and almost no entanglements form.
De Gennes\cite{degennes89} argued that the probability of entanglements across an immiscible interface at equilibrium should scale as $exp\left(-N_e\chi\right)$, reflecting the probability for a loop crossing the interface having length larger than $N_e$. This is qualitatively consistent with the loop statistics we measure (Figure 1 in the Supplemental Information) and explains the rapid drop in entanglements as $\tilde{\epsilon}_{12}$ decreases. 
Given $N_e \sim 85$ and our estimates of $\chi$,
$\exp\left(-N_e\chi\right)=0.08$, 0.02 and 0.0007 for $\epsilon=0.99$, 0.98 and 0.95, respectively.

There is a strong correlation between the threshold width for entanglement formation and interfacial shear strength.
Comparing \ref{fig:interface} and \ref{fig:strength}(b), we see that there is a sharp rise in $\sigma_{max}$ at the time when $\left<d\right>$ exceeds $1.5a$ for $\epsilon_{12}=1.0$ and 0.99.
Less miscible systems show little increase in strength because $\left<d\right>$ remains below the threshold value.
Experiments have also found that a minimum interfacial width is needed for the development of interfacial strength.\cite{schnell98,schnell99}.
One way of broadening interfaces in such immiscible systems is to add random copolymers, and experiments show this is effective in raising
interfacial strength\cite{creton02}. 
This will be an interesting topic for future simulation studies.

To summarize, we have demonstrated that the mechanical weakness of immiscible polymer interfaces is closely related to the lack of entanglements at the interface. The development of entanglements is greatly suppressed due to limited interdiffusion. At equilibrium, the density of entanglements is reduced compared to that in the bulk. Consequently, chains can be easily pulled out from the opposite side at a low stress. Our results also show that there is a minimum interdiffusion depth required for significant entanglement formation and therefore growth of the interfacial strength. These findings should help further development of theoretical descriptions of entanglement formation and fracture behavior at immiscible polymer-polymer interfaces, and also benefit engineering design of interfacial strengthening mechanisms. 
 
\begin{acknowledgement}
This work was supported in part by the National Science
Foundation under Grants No. DMR-1006805, No. CMMI-
0923018, and No. OCI-0963185.
M. O. R. acknowledges support from the Simons
Foundation. This research used resources at the National
Energy Research Scientific Computing Center, which is
supported by the Office of Science of the United States
Department of Energy under Contract No. DE-AC02-
05CH11231. Research was carried out in part at
the Center for Integrated Nanotechnologies, a U.S.
Department of Energy, Office of Basic Energy Sciences,
user facility. Sandia National Laboratories is a multiprogram
laboratory managed and operated by Sandia
Corporation, a wholly owned subsidiary of Lockheed
Martin Corporation, for the U.S. Department of Energy's
National Nuclear Security Administration under Contract
No. DE-AC04-94AL85000.

\end{acknowledgement}


{\bf SUPPORTING INFORMATION}



{\bf Model and Methodology Details}

All of the simulations presented here use the canonical bead-spring model \cite{kremer90} that captures the properties of linear homopolymers. Each polymer chain contains $N$ spherical beads of mass $m$. All beads interact via the truncated shifted Lennard-Jones potential
\begin{equation}
U_{\rm LJ} (r)=4u_0 [(a/r)^{12}-(a/r)^6-
(a/r_{\rm c} )^{12}+(a/r_{\rm c})^6] \ \ ,
\end{equation}
where $r_{\rm c}$ is the cutoff radius and $U_{\rm LJ} (r)=0$ for $r>r_{\rm c}$.
All quantities are expressed in terms of the molecular diameter $a$,
the interaction energy $u_0$, and the characteristic time $\tau=a(m/u_0 )^{1/2}$. 

For equilibration, beads along the chain were connected by an additional unbreakable finitely extensible nonlinear elastic (FENE) potential
\begin{equation}
U_{\rm FENE} (r)=-\frac{1}{2}kR_0^2 \ln [1-(r/R_0 )^2] \ \ ,
\end{equation}
with $R_0=1.5a$ and $k=30 u_0 a^{-2}$.
For mechanical tests, chain scission plays an essential role and a simple quartic potential was used
\begin{equation}
U_{Q} (r)=K(r-R_{\rm c} )^2 (r-R_{\rm c} )(r-R_{\rm c}-B)+U_0 \ \ ,
\end{equation}
with $K=2351u_0/k_B$, $B=-0.7425a$, $R_{\rm c}=1.5a$, and $U_0=92.74467u_0$. 

The equations of motion were integrated using a velocity-Verlet algorithm with a time step $\delta t$. The temperature was held constant by a Langevin thermostat with a damping constant $\Gamma$.\cite{kremer90} All simulations were carried out using the LAMMPS parallel MD code.\cite{plimpton95}

Two thin films were constructed following the standard methodology discussed by Auhl et al.~\cite{auhl03}. Each film contains $M=4800$ chains of length $N=500$ beads or a total of 2.4 million beads. Periodic boundary conditions were applied along the $x$- and $y$- directions with dimensions $L_x=700a$ and $L_y=40a$, while along the non-periodic $z$- direction $L_z=100a$. Each film was well equilibrated at a temperature $T=1.0 u_0/k_B$ with
$r_{\rm c}=2.5a$, $\Gamma = 0.1 \tau^{-1}$ and time step $\delta t=0.01\tau$. Pressure $P=0$ was maintained by expansion/contraction along the $x$-direction. For interdiffusion, the films were placed as close to contact as possible without overlap to form an interface at $z=0$. The volume was held fixed during the interdiffusion simulations by two repulsive walls perpendicular to the $z$-direction. 
After interdiffusion the density $\rho_I$ of monomers of each type $I$
was calculated as a function of height.
To remove the effect of long-wavelength capillary waves, the density was
averaged over $L_y$ and over regions of width $80a$ along $L_x$.
The local interface position was then identified
with the height where the two species had the same density.

For the mechanical test, we first reduced the cutoff radius to $r_{\rm c}=1.5a$ and the time step to $\delta t=0.005\tau$ to reduce density changes and facilitate comparison with past mechanical studies \cite{rottler02a, rottler02b, rottler03, hoy07, hoy08}. Then the temperature was quenched at constant volume with a rate $\dot{T} =-10^{-3}  u_0/(k_B \tau)$ to $T=0.5 u_0/k_B$ where $P=0$. Subsequent quenching to $T=0.2 u_0/k_B$ was done at $\dot{T} = -2 \times 10^{-4} u_0/(k_B \tau)$ and $P=0$. A Nose-Hoover barostat with time constant $50 \tau$ was applied to $P_{\rm xx}$ and $P_{\rm yy}$. The repulsive walls were maintained at $z=\pm L_z$. We verified that our conclusions are not sensitive to the details of the quench protocol or geometry.

In the shear test, beads within $5a$ of the top and bottom were held rigid and displaced at constant velocity
in opposite directions along the $y-$axis. The average strain rate in the film, $d\gamma/dt=2 \times 10^{-4} \tau^{-1}$, was low enough that it did not affect the mode of failure and stress had time to equilibrate across the system \cite{rottler03c}. The shear stress $\sigma$ was determined from the mean lateral force per unit area applied by the top and bottom walls. The temperature was maintained at $T=0.2 u_0/k_B$ with a Langevin thermostat ($\Gamma = 1 \tau^{-1}$) acting only on the x-component to avoid biasing the flow.

{\bf Interfacial Loop Statistics}

\begin{figure}[htb]
\includegraphics[width=0.45\textwidth]{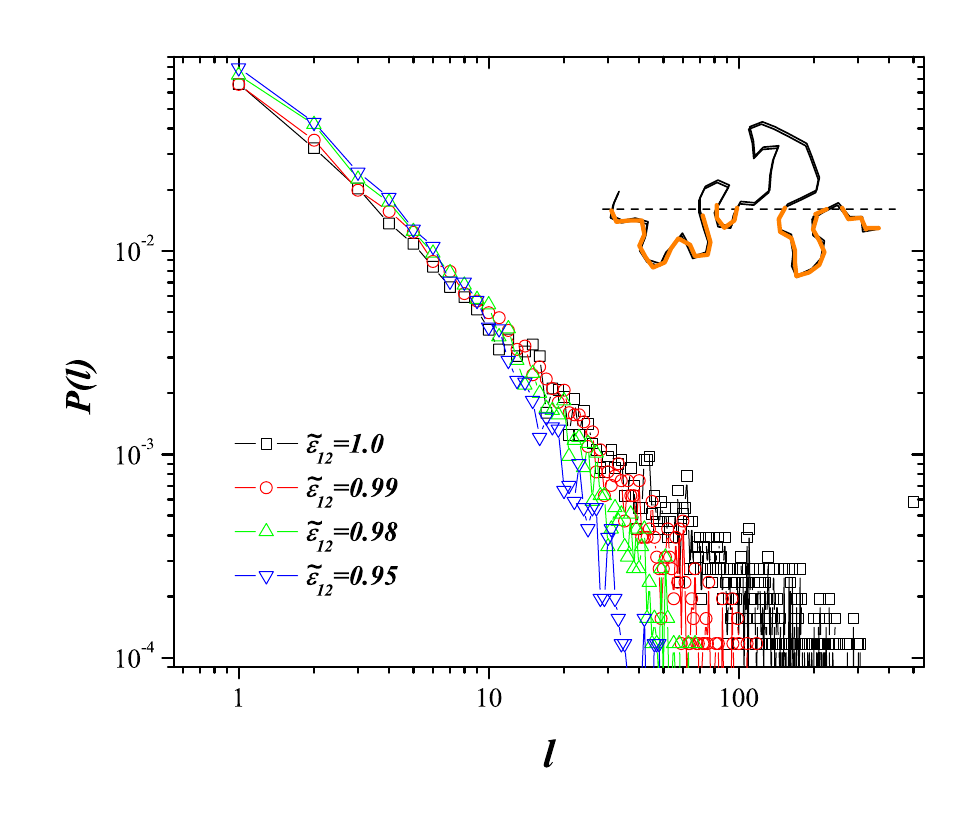}
\caption{Statistics of the length $l$ of interfacial loops.
}
\label{fig:loops}
\end{figure}
 
We define an interfacial loop as a segment of consecutive monomers penetrating into the other side, as schematically shown in \ref{fig:loops}(segments in orange color). For segments in the middle of the chain, both ends of the segment locate at the local mid-plane (dash line) across the interface. For segments including chain ends, only one end locates at the local mid-plane. The number of bonds in each segment is determined and denoted as $l$. The statistics $P(l)$ for the equilibrium immiscible interfaces and the miscible interface at $t=5M\tau$ are shown in \ref{fig:loops}.

The probability of long loops decays exponentially and the rate of decay increases with increasing immiscibility. As discussed in the main text, de Gennes\cite{degennes89} argued that interfacial entanglements should be limited to loops with $l$ greater than $N_e=85$. These are almost completely suppressed for $\tilde{\epsilon}_{12}=0.98$ and $0.95$ where $\left<d\right><d_c$. A few loops satisfy this criterion for $\tilde{\epsilon}_{12}=0.99$ and the number grows with time for $\tilde{\epsilon}_{12}=1.0$.
\bibliography{immiscible13}

\providecommand*\mcitethebibliography{\thebibliography}
\csname @ifundefined\endcsname{endmcitethebibliography}
  {\let\endmcitethebibliography\endthebibliography}{}
\begin{mcitethebibliography}{39}
\providecommand*\natexlab[1]{#1}
\providecommand*\mciteSetBstSublistMode[1]{}
\providecommand*\mciteSetBstMaxWidthForm[2]{}
\providecommand*\mciteBstWouldAddEndPuncttrue
  {\def\EndOfBibitem{\unskip.}}
\providecommand*\mciteBstWouldAddEndPunctfalse
  {\let\EndOfBibitem\relax}
\providecommand*\mciteSetBstMidEndSepPunct[3]{}
\providecommand*\mciteSetBstSublistLabelBeginEnd[3]{}
\providecommand*\EndOfBibitem{}
\mciteSetBstSublistMode{f}
\mciteSetBstMaxWidthForm{subitem}{(\alph{mcitesubitemcount})}
\mciteSetBstSublistLabelBeginEnd
  {\mcitemaxwidthsubitemform\space}
  {\relax}
  {\relax}

\bibitem[Helfand and Tagami(1971)Helfand, and Tagami]{helfand71}
Helfand,~E.; Tagami,~Y. \emph{J. Polym. Sci., Part B: Polym. Phys.}
  \textbf{1971}, \emph{9}, 741--746\relax
\mciteBstWouldAddEndPuncttrue
\mciteSetBstMidEndSepPunct{\mcitedefaultmidpunct}
{\mcitedefaultendpunct}{\mcitedefaultseppunct}\relax
\EndOfBibitem
\bibitem[Helfand and Tagami(1972)Helfand, and Tagami]{helfand72}
Helfand,~E.; Tagami,~Y. \emph{J. Chem. Phys.} \textbf{1972}, \emph{56},
  3592--3601\relax
\mciteBstWouldAddEndPuncttrue
\mciteSetBstMidEndSepPunct{\mcitedefaultmidpunct}
{\mcitedefaultendpunct}{\mcitedefaultseppunct}\relax
\EndOfBibitem
\bibitem[Rubinstein and Colby(2003)Rubinstein, and Colby]{rubinstein03}
Rubinstein,~M.; Colby,~R.~H. \emph{Polymer Physics}; OUP Oxford: Oxford,
  2003\relax
\mciteBstWouldAddEndPuncttrue
\mciteSetBstMidEndSepPunct{\mcitedefaultmidpunct}
{\mcitedefaultendpunct}{\mcitedefaultseppunct}\relax
\EndOfBibitem
\bibitem[Robeson(2007)]{robeson07}
Robeson,~L.~M. \emph{Polymer Blends: A Comprehensive Review}; Hanser-Gardner
  Publications: Cincinnati, OH, 2007\relax
\mciteBstWouldAddEndPuncttrue
\mciteSetBstMidEndSepPunct{\mcitedefaultmidpunct}
{\mcitedefaultendpunct}{\mcitedefaultseppunct}\relax
\EndOfBibitem
\bibitem[Wool(1995)]{wool95}
Wool,~R.~P. \emph{Polymer Interfaces: Structure and Strength}; Hanser: Munich,
  1995\relax
\mciteBstWouldAddEndPuncttrue
\mciteSetBstMidEndSepPunct{\mcitedefaultmidpunct}
{\mcitedefaultendpunct}{\mcitedefaultseppunct}\relax
\EndOfBibitem
\bibitem[Jones and Richards(1999)Jones, and Richards]{jone99}
Jones,~R. A.~L.; Richards,~R.~W. \emph{Polymers at Surfaces and Interfaces};
  Cambridge University Press: New York, 1999\relax
\mciteBstWouldAddEndPuncttrue
\mciteSetBstMidEndSepPunct{\mcitedefaultmidpunct}
{\mcitedefaultendpunct}{\mcitedefaultseppunct}\relax
\EndOfBibitem
\bibitem[Everaers et~al.(2004)Everaers, Sukumaran, Grest, Svaneborg,
  Sivasubramanian, and Kremer]{everaers04}
Everaers,~R.; Sukumaran,~S.~K.; Grest,~G.~S.; Svaneborg,~C.;
  Sivasubramanian,~A.; Kremer,~K. \emph{Science} \textbf{2004}, \emph{303},
  823--826\relax
\mciteBstWouldAddEndPuncttrue
\mciteSetBstMidEndSepPunct{\mcitedefaultmidpunct}
{\mcitedefaultendpunct}{\mcitedefaultseppunct}\relax
\EndOfBibitem
\bibitem[Kr\"oger(2005)]{kroger05}
Kr\"oger,~M. \emph{Comput. Phys. Commun.} \textbf{2005}, \emph{168},
  209--232\relax
\mciteBstWouldAddEndPuncttrue
\mciteSetBstMidEndSepPunct{\mcitedefaultmidpunct}
{\mcitedefaultendpunct}{\mcitedefaultseppunct}\relax
\EndOfBibitem
\bibitem[Tzoumanekas and Theodorou(2006)Tzoumanekas, and
  Theodorou]{tzoumanekas06}
Tzoumanekas,~C.; Theodorou,~D.~N. \emph{Macromolecules} \textbf{2006},
  \emph{39}, 4592--4604\relax
\mciteBstWouldAddEndPuncttrue
\mciteSetBstMidEndSepPunct{\mcitedefaultmidpunct}
{\mcitedefaultendpunct}{\mcitedefaultseppunct}\relax
\EndOfBibitem
\bibitem[de~Gennes(1971)]{degennes71}
de~Gennes,~P.~G. \emph{J. Chem. Phys.} \textbf{1971}, \emph{55}, 572--579\relax
\mciteBstWouldAddEndPuncttrue
\mciteSetBstMidEndSepPunct{\mcitedefaultmidpunct}
{\mcitedefaultendpunct}{\mcitedefaultseppunct}\relax
\EndOfBibitem
\bibitem[Doi and Edwards(1988)Doi, and Edwards]{doi88}
Doi,~M.; Edwards,~S.~F. \emph{The Theory of Polymer Dynamics}; Oxford
  University Press: Oxford, 1988\relax
\mciteBstWouldAddEndPuncttrue
\mciteSetBstMidEndSepPunct{\mcitedefaultmidpunct}
{\mcitedefaultendpunct}{\mcitedefaultseppunct}\relax
\EndOfBibitem
\bibitem[Schnell et~al.(1998)Schnell, Stamm, and Creton]{schnell98}
Schnell,~R.; Stamm,~M.; Creton,~C. \emph{Macromolecules} \textbf{1998},
  \emph{31}, 2284--2292\relax
\mciteBstWouldAddEndPuncttrue
\mciteSetBstMidEndSepPunct{\mcitedefaultmidpunct}
{\mcitedefaultendpunct}{\mcitedefaultseppunct}\relax
\EndOfBibitem
\bibitem[Schnell et~al.(1999)Schnell, Stamm, and Creton]{schnell99}
Schnell,~R.; Stamm,~M.; Creton,~C. \emph{Macromolecules} \textbf{1999},
  \emph{32}, 3420--3425\relax
\mciteBstWouldAddEndPuncttrue
\mciteSetBstMidEndSepPunct{\mcitedefaultmidpunct}
{\mcitedefaultendpunct}{\mcitedefaultseppunct}\relax
\EndOfBibitem
\bibitem[Brown(2001)]{brown01}
Brown,~H.~R. \emph{Macromolecules} \textbf{2001}, \emph{34}, 3720--3724\relax
\mciteBstWouldAddEndPuncttrue
\mciteSetBstMidEndSepPunct{\mcitedefaultmidpunct}
{\mcitedefaultendpunct}{\mcitedefaultseppunct}\relax
\EndOfBibitem
\bibitem[Creton et~al.(2001)Creton, Kramer, Brown, and C-Y.Hui]{creton02}
Creton,~C.; Kramer,~E.~J.; Brown,~H.~R.; C-Y.Hui, \emph{Adv. Polym. Sci.}
  \textbf{2001}, \emph{156}, 53--136\relax
\mciteBstWouldAddEndPuncttrue
\mciteSetBstMidEndSepPunct{\mcitedefaultmidpunct}
{\mcitedefaultendpunct}{\mcitedefaultseppunct}\relax
\EndOfBibitem
\bibitem[Cole et~al.(2003)Cole, Cook, and Macosko]{cole03}
Cole,~P.~J.; Cook,~R.~F.; Macosko,~C.~W. \emph{Macromolecules} \textbf{2003},
  \emph{36}, 2808--2815\relax
\mciteBstWouldAddEndPuncttrue
\mciteSetBstMidEndSepPunct{\mcitedefaultmidpunct}
{\mcitedefaultendpunct}{\mcitedefaultseppunct}\relax
\EndOfBibitem
\bibitem[Boiko(2012)]{boiko12b}
Boiko,~Y.~M. \emph{Colloid Polym. Sci.} \textbf{2012}, \emph{290},
  1201--1206\relax
\mciteBstWouldAddEndPuncttrue
\mciteSetBstMidEndSepPunct{\mcitedefaultmidpunct}
{\mcitedefaultendpunct}{\mcitedefaultseppunct}\relax
\EndOfBibitem
\bibitem[de~Gennes(1989)]{degennes89}
de~Gennes,~P.~G. \emph{C. R. Acad. Sci. Ser. II} \textbf{1989}, \emph{308},
  1401--1403\relax
\mciteBstWouldAddEndPuncttrue
\mciteSetBstMidEndSepPunct{\mcitedefaultmidpunct}
{\mcitedefaultendpunct}{\mcitedefaultseppunct}\relax
\EndOfBibitem
\bibitem[Benkoski et~al.(2002)Benkoski, Fredrickson, and Kramer]{benkoski02}
Benkoski,~J.~J.; Fredrickson,~G.~H.; Kramer,~E.~J. \emph{J. Polym. Sci., Part
  B: Polym. Phys.} \textbf{2002}, \emph{40}, 2377--2386\relax
\mciteBstWouldAddEndPuncttrue
\mciteSetBstMidEndSepPunct{\mcitedefaultmidpunct}
{\mcitedefaultendpunct}{\mcitedefaultseppunct}\relax
\EndOfBibitem
\bibitem[Silvestri et~al.(2003)Silvestri, Brown, Carra, and Carra]{silvestri03}
Silvestri,~L.; Brown,~H.~R.; Carra,~S.; Carra,~S. \emph{J. Chem. Phys.}
  \textbf{2003}, \emph{119}, 8140--8149\relax
\mciteBstWouldAddEndPuncttrue
\mciteSetBstMidEndSepPunct{\mcitedefaultmidpunct}
{\mcitedefaultendpunct}{\mcitedefaultseppunct}\relax
\EndOfBibitem
\bibitem[Hoy and Grest(2007)Hoy, and Grest]{hoy07b}
Hoy,~R.~S.; Grest,~G.~S. \emph{Macromolecules} \textbf{2007}, \emph{40},
  8389--8395\relax
\mciteBstWouldAddEndPuncttrue
\mciteSetBstMidEndSepPunct{\mcitedefaultmidpunct}
{\mcitedefaultendpunct}{\mcitedefaultseppunct}\relax
\EndOfBibitem
\bibitem[Kremer and Grest(1990)Kremer, and Grest]{kremer90}
Kremer,~K.; Grest,~G.~S. \emph{J. Chem. Phys.} \textbf{1990}, \emph{92},
  5057--5086\relax
\mciteBstWouldAddEndPuncttrue
\mciteSetBstMidEndSepPunct{\mcitedefaultmidpunct}
{\mcitedefaultendpunct}{\mcitedefaultseppunct}\relax
\EndOfBibitem
\bibitem[Rottler et~al.(2002)Rottler, Barsky, and Robbins]{rottler02a}
Rottler,~J.; Barsky,~S.; Robbins,~M.~O. \emph{Phys. Rev. Lett.} \textbf{2002},
  \emph{89}, 148304\relax
\mciteBstWouldAddEndPuncttrue
\mciteSetBstMidEndSepPunct{\mcitedefaultmidpunct}
{\mcitedefaultendpunct}{\mcitedefaultseppunct}\relax
\EndOfBibitem
\bibitem[Stevens(2001)]{stevens01}
Stevens,~M.~J. \emph{Macromolecules} \textbf{2001}, \emph{34}, 2710--2718\relax
\mciteBstWouldAddEndPuncttrue
\mciteSetBstMidEndSepPunct{\mcitedefaultmidpunct}
{\mcitedefaultendpunct}{\mcitedefaultseppunct}\relax
\EndOfBibitem
\bibitem[Ge et~al.(2013)Ge, Pierce, Perahia, Grest, and Robbins]{ge13}
Ge,~T.; Pierce,~F.; Perahia,~D.; Grest,~G.~S.; Robbins,~M.~O. \emph{Phys. Rev.
  Lett.} \textbf{2013}, \emph{110}, 98301\relax
\mciteBstWouldAddEndPuncttrue
\mciteSetBstMidEndSepPunct{\mcitedefaultmidpunct}
{\mcitedefaultendpunct}{\mcitedefaultseppunct}\relax
\EndOfBibitem
\bibitem[Rottler and Robbins(2002)Rottler, and Robbins]{rottler02b}
Rottler,~J.; Robbins,~M.~O. \emph{Phys. Rev. Lett.} \textbf{2002}, \emph{89},
  195501\relax
\mciteBstWouldAddEndPuncttrue
\mciteSetBstMidEndSepPunct{\mcitedefaultmidpunct}
{\mcitedefaultendpunct}{\mcitedefaultseppunct}\relax
\EndOfBibitem
\bibitem[Rottler and Robbins(2003)Rottler, and Robbins]{rottler03}
Rottler,~J.; Robbins,~M.~O. \emph{Phys. Rev. E} \textbf{2003}, \emph{68},
  011801\relax
\mciteBstWouldAddEndPuncttrue
\mciteSetBstMidEndSepPunct{\mcitedefaultmidpunct}
{\mcitedefaultendpunct}{\mcitedefaultseppunct}\relax
\EndOfBibitem
\bibitem[Hoy and Robbins(2007)Hoy, and Robbins]{hoy07}
Hoy,~R.~S.; Robbins,~M.~O. \emph{Phys. Rev. Lett.} \textbf{2007}, \emph{99},
  117801\relax
\mciteBstWouldAddEndPuncttrue
\mciteSetBstMidEndSepPunct{\mcitedefaultmidpunct}
{\mcitedefaultendpunct}{\mcitedefaultseppunct}\relax
\EndOfBibitem
\bibitem[Hoy and Robbins(2008)Hoy, and Robbins]{hoy08}
Hoy,~R.~S.; Robbins,~M.~O. \emph{Phys. Rev. E} \textbf{2008}, \emph{77},
  031801\relax
\mciteBstWouldAddEndPuncttrue
\mciteSetBstMidEndSepPunct{\mcitedefaultmidpunct}
{\mcitedefaultendpunct}{\mcitedefaultseppunct}\relax
\EndOfBibitem
\bibitem[Rottler and Robbins(2003)Rottler, and Robbins]{rottler03c}
Rottler,~J.; Robbins,~M.~O. \emph{Phys. Rev. E} \textbf{2003}, \emph{68},
  011507\relax
\mciteBstWouldAddEndPuncttrue
\mciteSetBstMidEndSepPunct{\mcitedefaultmidpunct}
{\mcitedefaultendpunct}{\mcitedefaultseppunct}\relax
\EndOfBibitem
\bibitem[Pierce et~al.(2011)Pierce, Perahia, and Grest]{pierce11}
Pierce,~F.; Perahia,~D.; Grest,~G.~S. \emph{Europhys. Lett.} \textbf{2011},
  \emph{95}, 46001\relax
\mciteBstWouldAddEndPuncttrue
\mciteSetBstMidEndSepPunct{\mcitedefaultmidpunct}
{\mcitedefaultendpunct}{\mcitedefaultseppunct}\relax
\EndOfBibitem
\bibitem[ge1()]{ge13b}
Ge, T.; Robbins, M. O.; Perahia, D.; Grest, G. S. manuscript in
  preparation.\relax
\mciteBstWouldAddEndPunctfalse
\mciteSetBstMidEndSepPunct{\mcitedefaultmidpunct}
{}{\mcitedefaultseppunct}\relax
\EndOfBibitem
\bibitem[Lacasse et~al.(1998)Lacasse, Grest, and Levine]{lacasse98}
Lacasse,~M.-D.; Grest,~G.~S.; Levine,~A.~J. \emph{Phys. Rev. Lett.}
  \textbf{1998}, \emph{80}, 309\relax
\mciteBstWouldAddEndPuncttrue
\mciteSetBstMidEndSepPunct{\mcitedefaultmidpunct}
{\mcitedefaultendpunct}{\mcitedefaultseppunct}\relax
\EndOfBibitem
\bibitem[Silberberg(1982)]{silberberg82}
Silberberg,~A. \emph{J. Colloid Interface Sci.} \textbf{1982}, \emph{90},
  86--91\relax
\mciteBstWouldAddEndPuncttrue
\mciteSetBstMidEndSepPunct{\mcitedefaultmidpunct}
{\mcitedefaultendpunct}{\mcitedefaultseppunct}\relax
\EndOfBibitem
\bibitem[Theodorou(1988)]{theodorou88}
Theodorou,~D.~N. \emph{Macromolecules} \textbf{1988}, \emph{21},
  1400--1410\relax
\mciteBstWouldAddEndPuncttrue
\mciteSetBstMidEndSepPunct{\mcitedefaultmidpunct}
{\mcitedefaultendpunct}{\mcitedefaultseppunct}\relax
\EndOfBibitem
\bibitem[Silberberg(1988)]{silberberg88}
Silberberg,~A. \emph{J. Colloid Interface Sci.} \textbf{1988}, \emph{125},
  14--22\relax
\mciteBstWouldAddEndPuncttrue
\mciteSetBstMidEndSepPunct{\mcitedefaultmidpunct}
{\mcitedefaultendpunct}{\mcitedefaultseppunct}\relax
\EndOfBibitem
\bibitem[Plimpton(1995)]{plimpton95}
Plimpton,~S. \emph{J. Comp. Phys.} \textbf{1995}, \emph{117}, 1--19\relax
\mciteBstWouldAddEndPuncttrue
\mciteSetBstMidEndSepPunct{\mcitedefaultmidpunct}
{\mcitedefaultendpunct}{\mcitedefaultseppunct}\relax
\EndOfBibitem
\bibitem[Auhl et~al.(2003)Auhl, Everaers, Grest, Kremer, and Plimpton]{auhl03}
Auhl,~R.; Everaers,~R.; Grest,~G.~S.; Kremer,~K.; Plimpton,~S.~J. \emph{J.
  Chem. Phys.} \textbf{2003}, \emph{119}, 12718--12728\relax
\mciteBstWouldAddEndPuncttrue
\mciteSetBstMidEndSepPunct{\mcitedefaultmidpunct}
{\mcitedefaultendpunct}{\mcitedefaultseppunct}\relax
\EndOfBibitem
\end{mcitethebibliography}

\end{document}